\documentclass[aps, pre, twocolumn, superscriptaddress, nofootinbib]{revtex4-1}
\usepackage[utf8]{inputenc}
\usepackage{graphicx}
\usepackage{xcolor}
\usepackage{amsmath}

\definecolor{darkblue}{rgb}{0,0,0.6}
\definecolor{darkred}{rgb}{0.6,0,0}
\usepackage[colorlinks=true, urlcolor=darkblue, citecolor=darkblue, linkcolor=darkred]{hyperref}

\begin{document}

\title{Cylinder morphology of a stretched and twisted ribbon}
\author{Vincent D\'emery}
\affiliation{Gulliver, CNRS, ESPCI Paris, PSL Research University, 10 rue Vauquelin, Paris, France}
\affiliation{Univ Lyon, ENS de Lyon, Univ Claude Bernard Lyon 1, CNRS, Laboratoire de Physique, F-69342 Lyon, France}
\author{Huy Pham Dinh}
\affiliation{Laboratoire Interfaces $\&$ Fluides Complexes, Universit\'e de Mons, 20 Place du Parc, B-7000 Mons, Belgium.}
\author{Pascal Damman}
\affiliation{Laboratoire Interfaces $\&$ Fluides Complexes, Universit\'e de Mons, 20 Place du Parc, B-7000 Mons, Belgium.}

\date{\today}

\begin{abstract}
A rich zoology of shapes emerges from a simple stretched and twisted elastic ribbon.
Despite a lot of interest, all these shape are not understood, in particular the shape that prevails at large tension and twist and that emerges from a transverse instability of the helicoid.
Here, we propose a simple description for this cylindrical shape.
By comparing its energy to the energy of other configurations, we are able to determine its location on the phase diagram.
The theoretical predictions are in good agreement with our experimental results.
\end{abstract}
\maketitle

\section{Introduction}\label{}

Thin elastic sheets exhibit a wide variety of patterns in response to external loadings such as wrinkles, crumples, and folds~\cite{BenAmar1997, Witten2007, Pocivavsek2008, King2012, Paulsen2015, Paulsen2017}.
This rich behavior stems from their two dimensional nature, which introduces a coupling between mechanics and geometry.
In this context, a stretched and twisted ribbon is a remarkable playground: varying two parameters, the tension and the twist, allows to produce many different shapes, which can be organized on a phase diagram~\cite{Chopin2013}.
Understanding the emergence of these shapes is a considerable theoretical challenge.

\begin{figure}
\begin{center}
\includegraphics[width=\linewidth]{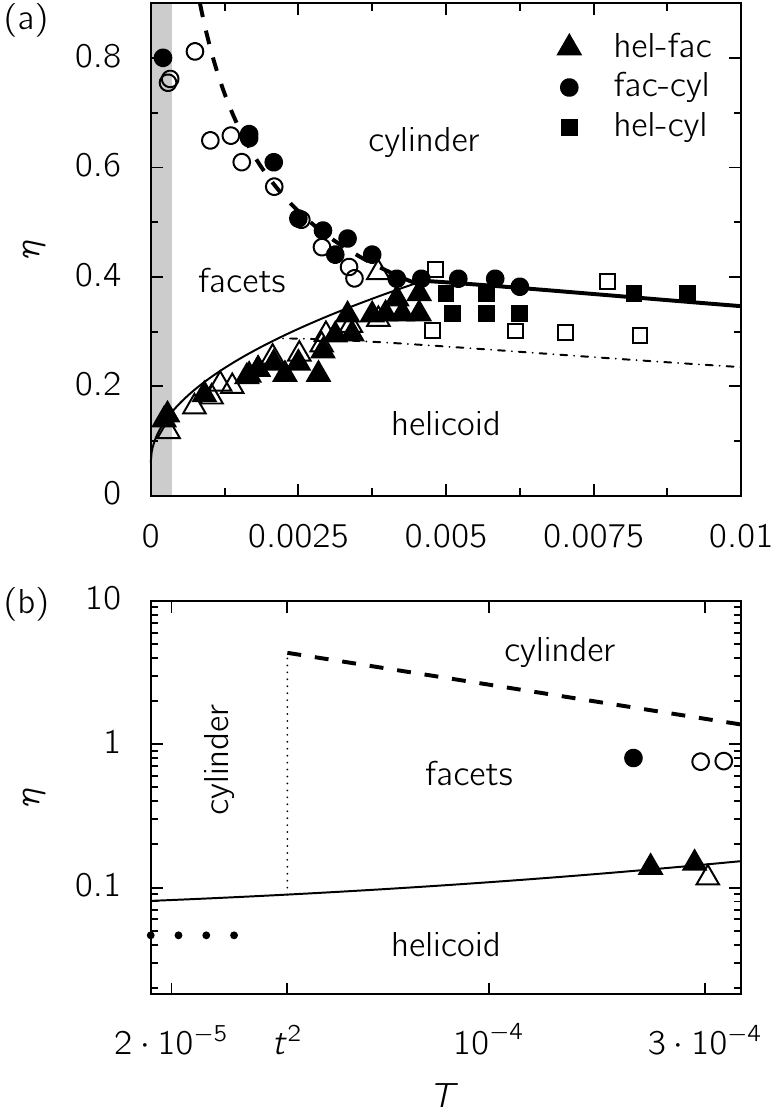}
\end{center}
\caption{Phase diagram with transitions between helicoid, cylinder and facetted shapes at (a) moderate tension and (b) low tension (expansion of the grey domain in (a)).
Experimental data for $t=6\cdot 10^{-3}$ (filled symbols, PET ribbons), data of Ref.~\cite{Chopin2013} (open symbols) and theoretical curves. 
Thin solid line, longitudinal linear instability of the helicoid; 
dashed-dotted line, transverse linear instability;
thick solid line, helicoid-cylinder transition from energy comparison;
dashed line, helicoid-facets with isometric ridges transition from energy comparison;
thick dotted line, helicoid-cylinder transition at low tension from the energy comparison;
thin dotted line, low tension limit of definition of the facets with isometric ridges.
}
\label{fig:phase_diag}
\end{figure}

The phase diagram of the stretched and twisted ribbon can be seen as organized around the helicoid~\cite{Chopin2013, Chopin2015}, which may become unstable and give birth to different, more complex shapes (Fig.~\ref{fig:phase_diag}(a)).
The first instability that has been understood is the longitudinal buckling instability: as the helicoid is twisted at relatively low tension, the center line is under compression and eventually buckles, forming wrinkles~\cite{Green1936, Green1937, Coman2008}.
Far from threshold, a facetted morphology is observed~\cite{Chopin2013}, which can be described as flat facets connected by isometric or stretching ridges~\cite{Bohr2013, PhamDinh2016}.
At very low tension, it has been suggested that a cylindrical wrapping may be the energetically favored state upon increasing the twist~\cite{Chopin2015}.
Finally, twisting at large tension, the helicoid undergoes a transverse buckling instability~\cite{Chopin2015}, which finally leads to self-contact~\cite{Chopin2013}.
These two unstabilities meet at a triple point, the so-called ``$\lambda$-point''~\cite{Chopin2015}.

\begin{figure}
\begin{center}
\includegraphics[width=0.9\linewidth]{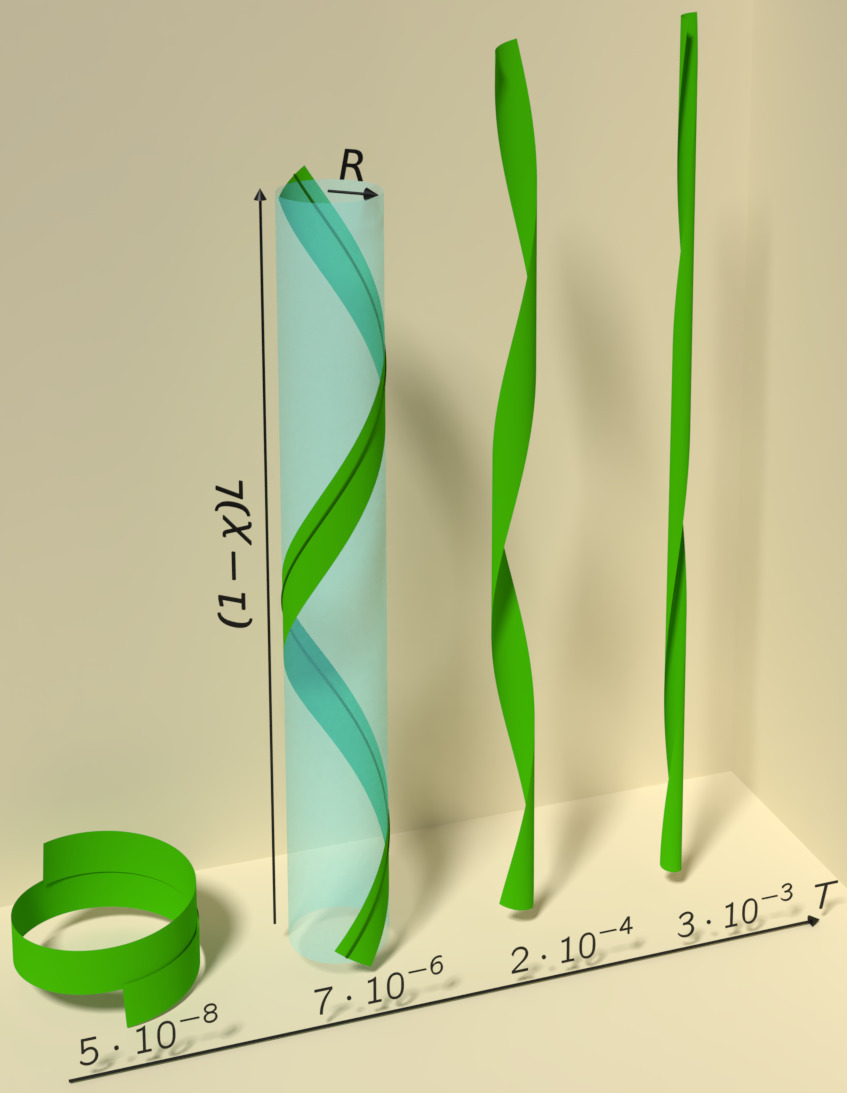}
\end{center}
\caption{Cylindrical configurations determined theoretically for $\eta=0.5$, $t=6\cdot 10^{-3}$, and different values of the tension $T$.
The parameters of the cylinder, the radius $R$ and the contraction $\chi$, are represented for $T=7\cdot 10^{-6}$.
The dark green line represents a line parallel to the centerline of the ribbon.}
\label{fig:cylinders}
\end{figure}

While the facetted morphology arising beyond the longitudinal instability has been studied on its own~\cite{Bohr2013, PhamDinh2016}, the ultimate shape of the ribbon after the transverse buckling instability remains unexplored.
Here, we propose an ansatz for this shape, whereby the ribbon wraps around a cylinder (Fig.~\ref{fig:cylinders}).
We compute the parameters of the cylinder over the whole phase diagram, and determine when it should be observed by comparing its energy to those of previously known shapes: the helicoid at vanishing and high tension, and the facetted shape with isometric ridges at intermediate tension.
We find a good agreement between the theoretical predictions, the experimental phase diagram~\cite{Chopin2013}, our experiments at large tension.
We also get a good agreement with previous theoretical predictions based on linear stability analysis of the helicoid~\cite{Coman2008, Chopin2015}.
Moreover, from this comparison we get insights into the nature of the helicoid-cylinder transition at very low tension on one hand, and at high tension on the other hand.

This article is organized as follows.
After the experimental method described in Sec.~\ref{sec:experiment}, we define the cylinder ansatz in Sec.~\ref{sec:cylinder}. 
Its main properties are computed for any values of the tension and the twist; the energy can be found numerically in general, and analytical expressions are given in limiting cases.
In Sec.~\ref{sec:energy_comparison}, we compare the energy of the cylinder to the energy of other shapes, depending on the tension, and obtain the region of existence of the cylinder in the phase diagram. 
For the sake of clarity, comparisons between theory and experiments were achieved through all the article.
We conclude in Sec.~\ref{sec:conclusion}.

\section{Experimental setup}\label{sec:experiment}

We use ribbons of length $L$, width $W$, and thickness $t$ under an external tension $F$ and clamped at their short edges, which are twisted relative to each other by a prescribed angle $\theta$.
Our ribbons are composed of polyethylene terephthalate PET (Young's modulus $E \simeq 3\,\textrm{GPa}$). 
The twisting and stretching of a ribbon are characterized by the dimensionless twist, $\eta=W\theta/L$, and dimensionless tension, $T=F/(WY)$, where $Y=tE$ is the stretching modulus. 
Following Refs.~\cite{Chopin2015, PhamDinh2016}, we use the width $W$ of the ribbon as the unit of length and the stretching modulus $Y$ as the unit of in-plane stress. 
All energies are given per unit length.
For twisting experiments at constant length, the tension is measured with a ProbeTack device equipped with a motorized translation system and a force transducer.

\section{Cylinder: definition and properties}\label{sec:cylinder}

\subsection{Definition}\label{}

We define the cylinder as a configuration where each line parallel to the midline of the ribbon winds around a cylinder of radius $R$ at a rate $\eta$, with a difference in height between the two ends of the ribbon given by $(1-\chi)L$, where $\chi$ is the contraction (Fig. \ref{fig:cylinders}).
If the length of the ribbon, or equivalently the contraction $\chi$, is fixed, the radius $R$ should be determined via energy minimization.
If instead the tension is imposed, both the radius and the contraction should be determined via energy minimization.

\subsection{Elastic energy}

The elastic energy of the ribbon is the sum of the stretching energy and the bending energy.
Both energies are easy to compute, because the strain and curvature are uniform in the ribbon.
The strain of the line parallel to the centerline of the ribbon is given by the Pythagorean theorem: 
\begin{equation}\label{eq:strain}
\epsilon=\sqrt{(1-\chi)^2+\eta^2R^2}-1.
\end{equation}
The curvature is $1/R$, so that the bending energy is $B/(2R^2)$, where $B$ is the bending modulus, given by $B=t^2/[12(1-\nu^2)]$ where $t, \nu$  are the thickness and the Poisson's ratio.
In order the keep track of the thickness in the following, but without getting heavy expressions, we set the Poisson's ratio to 0 so that $B=t^2/12$. The dependence on the Poisson ratio can be restored by replacing $t^2$ by $t^2/(1-\nu^2)$.
Finally, the elastic energy of the ribbon reads
\begin{equation}\label{eq:elastic_energy}
U^\text{el}_\text{cyl} = \frac{\epsilon^2}{2}+\frac{t^2}{24 R^2}.
\end{equation}
With this elastic energy, we can compute the free parameters of the ribbon in the fixed length and fixed tension configurations.

\subsection{Fixed length}\label{sub:fixed_length}

We restrict the study of the cylinder at fixed length to the case where the  length is fixed to its value at rest, $\chi=0$, which we investigated experimentally.
The strain is obtained from Eq.~(\ref{eq:strain}) with $\chi=0$: $\epsilon=\sqrt{1+\eta^2R^2}-1$.
Considering small strains, $\eta R$ should be small and we can approximate $\epsilon\simeq \eta^2 R^2/2$.
The elastic energy thus becomes,
\begin{equation}
U^\text{el}_\text{cyl} = \frac{\eta^4 R^4}{8}+\frac{t^2}{24R^2}.
\end{equation}

\begin{figure}
\begin{center}
\includegraphics[]{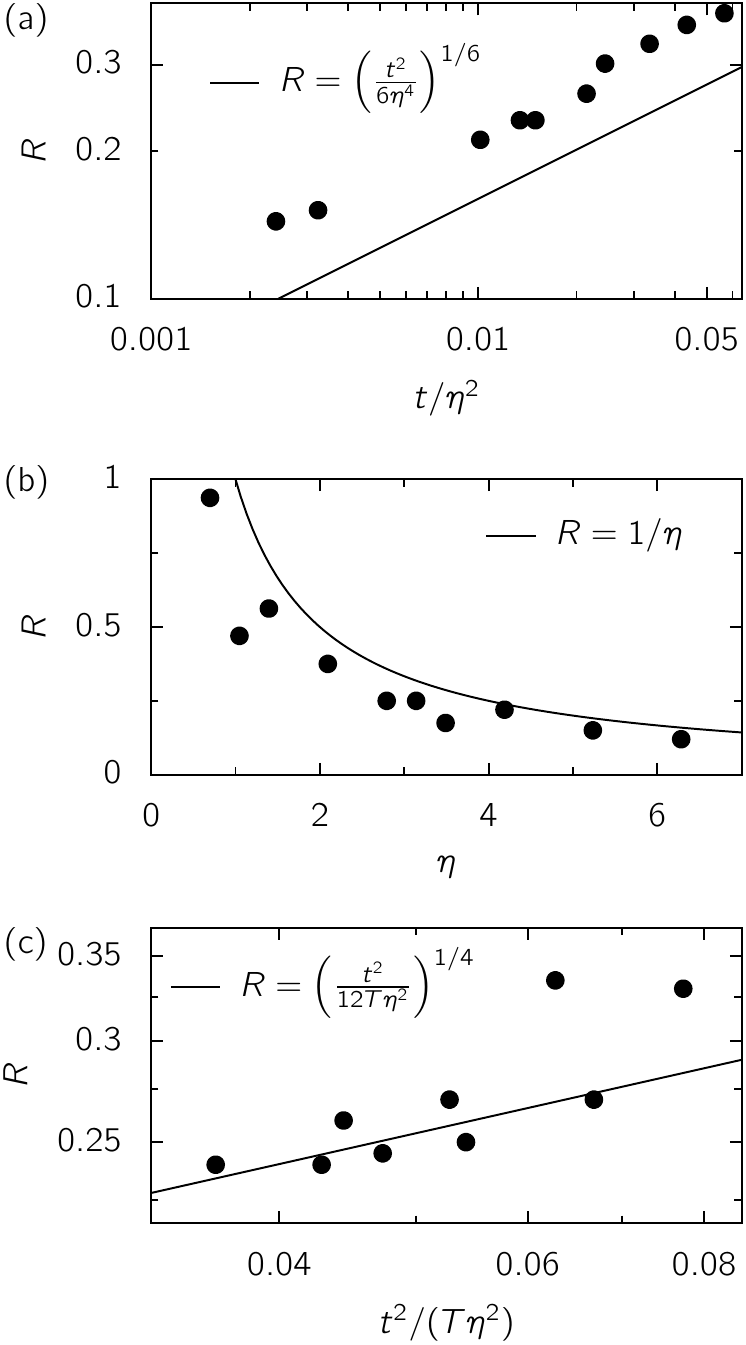}
\end{center}
\caption{Evolution of the measured cylinder radius for twisted PET ribbons ($t\simeq 6\cdot 10^{-3}$) in the three regimes : (a) Fixed length, (b) vanishing tension and (c) moderate tension. The solid lines correspond to the values predicted by the theoretical model. There is no fitting parameters.}
\label{fig:radius}
\end{figure}

The radius $R$ is set by the minimization of the elastic energy, which leads to
\begin{equation}\label{eq:cyl_radius_chi0}
R = \left(\frac{t^2}{6\eta^4} \right)^{1/6}.
\end{equation}
This relation agrees well with the experimental measurements (Fig.~\ref{fig:radius}(a)); the shift can be attributed to the finite length of the ribbon, which is not taken into account in the theory.
The elastic energy then becomes,
\begin{equation}\label{eq:cyl_energy_chi0}
U_\text{cyl}^\mathrm{el}= \frac{3}{8\times 6^{2/3}} (t\eta)^{4/3}\simeq 0.11 (t\eta)^{4/3}.
\end{equation}

\begin{figure}
\begin{center}
\includegraphics[]{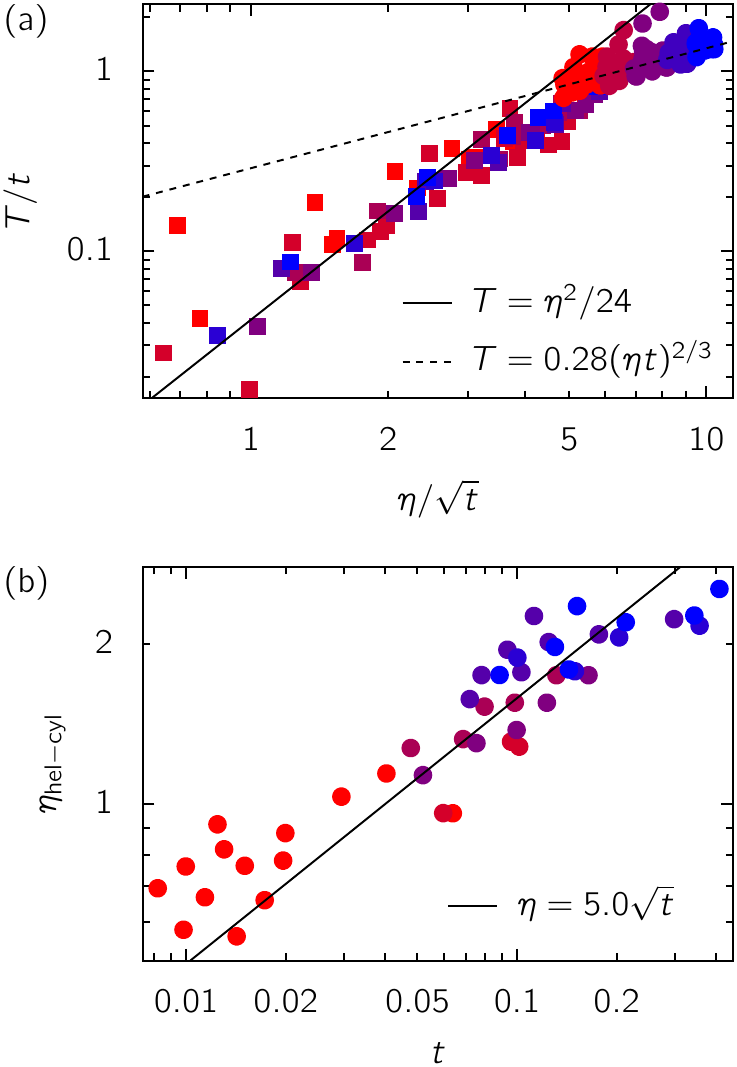}
\end{center}
\caption{(a) Evolution of the tension with $\eta$ at constant length showing the helicoid-cylinder transition ($T=\eta^2/24$ for the helicoid and $T \simeq 0.28(\eta t)^{2/3}$ for the cylinder).
(b)Transition helicoid-Cylinder at fixed length for various ribbon thicknesses.}
\label{fig:eta_helcyl_chi0}
\end{figure}

The tension in the ribbon is the variable conjugate to the contraction $\chi$, it is given by 
\begin{equation}\label{eq:def_tension}
T = -\frac{\partial U_\text{cyl}^\mathrm{el}}{\partial\chi}.
\end{equation}
It can be shown without approximation (App.~\ref{ap:tension_R_chi}) that it is given by
\begin{equation}
T = \frac{t^2}{12\eta^2 R^4}(1-\chi)
\end{equation}
for the radius $R$ that minimizes the elastic energy.
For $\chi=0$ and the radius $R$ given by Eq.~(\ref{eq:cyl_radius_chi0}), we find
\begin{equation}\label{eq:tension_const_length}
T = \frac{6^{2/3}}{12}(\eta t)^{2/3}\simeq 0.28(\eta t)^{2/3}.
\end{equation}
We note that this expression is equal to the longitudinal strain, $\epsilon\simeq \eta^2 R^2/2$.
This relation also quantitatively agrees with the tension measured in experiments without any fitting parameter (Fig.~\ref{fig:eta_helcyl_chi0}(a)).

\subsection{Fixed tension}\label{}

When a tension $T$ is imposed, the energy that should be minimized is the sum of the elastic energy, Eq.~(\ref{eq:elastic_energy}), and the ``potential energy'' $T\chi$~\cite{Chopin2015}:
\begin{equation}\label{eq:cyl_tot_energy}
U_\text{cyl}=U^\text{el}_\text{cyl} + T\chi= \frac{\epsilon^2}{2}+\frac{t^2}{24 R^2}+T\chi;
\end{equation}
this operation is a Legendre transform, since $T$ and $\chi$ are conjugate variables.
The contraction $\chi$ and the radius of the cylinder $R$ should be obtained by minimizing the total energy, Eq.~(\ref{eq:cyl_tot_energy}).

\begin{figure}
\begin{center}
\includegraphics[]{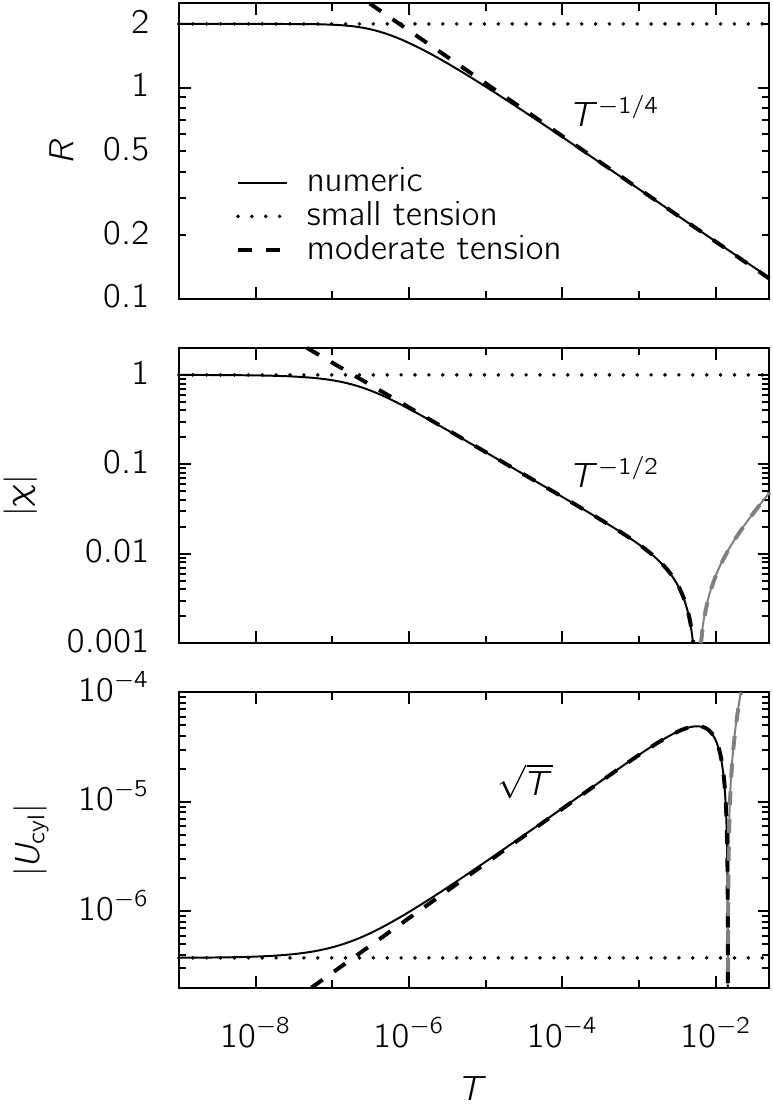}
\end{center}
\caption{
Radius $R$, contraction $\chi$, and energy $U_\text{cyl}$ of a cylinder as a function of the tension $T$ for $t=6\cdot 10^{-3}$ and $\eta=0.5$.
The numerical minimization (solid lines) of the energy is compared to the analytical expressions obtained in the small tension (dotted lines, $R=1/\eta,\ \chi=1,\ U_\text{cyl}=\eta^2t^2/24$) and moderate tension (dashed lines) regimes.
When values of the contraction or the energy are negative, their absolute value is plotted in gray.
The scaling laws for $(\eta t)^2\ll T\ll (\eta t)^{2/3}$ are indicated.
}
\label{fig:R_chi_U_approx}
\end{figure}

The minimization can be performed numerically (Fig.~\ref{fig:R_chi_U_approx}).
In order to get analytical expressions, we consider two limiting cases: the low tension regime, where the tension term can be neglected in the energy, Eq.~(\ref{eq:cyl_tot_energy}), and the moderate tension regime, where the contraction is small, $|\chi|\ll 1$.
The location of these regimes in the phase diagram is discussed below.

At low tension, we neglect the contraction term $T\chi$ in the energy, Eq.~(\ref{eq:cyl_tot_energy}). 
Minimizing the energy with respect to the contraction leads to $\epsilon \frac{\partial\epsilon}{\partial\chi}=0$, hence $\epsilon=0$ or $\partial\epsilon/\partial\chi=0$.
If $\epsilon=0$, the contraction is $\chi=1-\sqrt{1-\eta^2R^2}$, which imposes $\eta R\leq 1$; minimizing then the bending energy with respect to the radius amounts to maximize the radius, setting $R=1/\eta$, which in turn leads to $\chi=1$ and $U=\eta^2 t^2/24$.
If $\partial\epsilon/\partial\chi=0$, $\chi=1$ and $\epsilon=\eta R-1$, and the energy is $U=(\eta R-1)^2/2+t^2/(24R^2)$; minimizing the energy with respect to $R$ leads to a lower energy than the first condition (which leads to a particular case), and $R=(1+\alpha)/\eta$, where $\alpha$ satisfies $\alpha=\eta^2t^2/[12(1+\alpha)^3]$.
Since $t^2\ll 1$, $\alpha\sim t^2\ll 1$ and $R\simeq 1/\eta$; at the lowest order in $t$, the energy is
\begin{equation}\label{eq:energy_cyl_low_T}
U_\text{cyl}^{T=0} = \frac{\eta^2 t^2}{24}.
\end{equation}
The cylinder radius measured for vanishing tension agrees well with the proposed relation $R = 1/\eta$ (Fig.~\ref{fig:radius}(b)).

At moderate tension, we assume that the contraction is small; since the strain is also small, we can make the approximation
\begin{equation}
\epsilon=\sqrt{(1-\chi)^2+\eta^2R^2}-1\simeq \frac{\eta^2 R^2}{2} - \chi.
\end{equation}
Using this expression, the total energy (Eq.~(\ref{eq:cyl_tot_energy})) reads
\begin{equation}
U_\text{cyl}(\eta,\chi,R)=\frac{1}{2}\left(\frac{\eta^2R^2}{2}-\chi \right)^2 +\frac{t^2}{24R^2} +\chi T.
\end{equation}
Minimizing this expression first with respect to the contraction, and then with respect to the radius gives
\begin{align}
\chi & = \frac{\eta^2 R^2}{2} - T = \frac{\eta t}{4\sqrt{3}\sqrt{T}}-T, \label{eq:chi_moderate_tension}\\
R & = \left(\frac{t^2}{12 T\eta^2} \right)^{1/4}, \label{eq:r_moderate_tension}\\
U_\text{cyl} & = \frac{t\eta\sqrt{T}}{2\sqrt{3}}-\frac{T^2}{2}. \label{eq:energy_moderate_tension}
\end{align}
Fig.~\ref{fig:radius}(c) shows the good quantitative agreement between measured radius at moderate tension and Eq.~(\ref{eq:r_moderate_tension}), without any fitting parameter.
We have assumed that the contraction is small. From the expression of the contraction (Eq.~(\ref{eq:chi_moderate_tension})), we see that this assumption holds as long as $\eta^2 t^2\ll T\ll 1$.
Since the tension should be small to remain in the linear elasticity regime, and the twist can be of order 1, the moderate tension regime corresponds to $T\gg t^2$.

Two regimes can be identified in the moderate tension regime, from the expressions of the contraction and the energy, Eqs.~(\ref{eq:chi_moderate_tension}, \ref{eq:energy_moderate_tension}), which change sign for $T\sim(\eta t)^{2/3}$.
The regime where $T\gg (\eta t)^{2/3}$ corresponds to a ribbon under pure tension, as the contraction and energy are independent of the twist, $\chi= -T$ and $U=-T^2/2$.

The analytical expressions for the parameters and the energy of the cylinder in the different regimes are summarized in Table~\ref{tab:R_chi_U_approx}, and they are compared to the numerical minimization of the energy in Fig.~\ref{fig:R_chi_U_approx}.

\begin{table}
{ 
\renewcommand{\arraystretch}{1.5}
\begin{center}
\begin{tabular}{c||c|c|c}
regime & low tension & moderate tension & fixed length $\chi=0$  \\ \hline \hline
$T$ & $T\ll t^2$ & $T\gg t^2$ & $T = \frac{6^{2/3}}{12}(\eta t)^{2/3}$  \\ \hline
$\chi$ & $1$ & $\frac{\eta t}{4\sqrt{3}\sqrt{T}}- T$& $0$ \\ \hline
$R$ & $1/\eta$ & $\left(\frac{t^2}{12T\eta^2} \right)^{1/4}$& $\left(\frac{t^2}{6\eta^4} \right)^{1/6}$ \\ \hline
$U_\text{cyl}$ & $\frac{\eta^2 t^2}{24}$ & $\frac{t\eta\sqrt{T}}{2\sqrt{3}}-\frac{T^2}{2}$& $\frac{3}{8\times 6^{2/3}} (t\eta)^{4/3}$ 
\end{tabular}
\end{center}
}

\caption{Expressions for the radius $R$, the contraction $\chi$ and the energy $U_\text{cyl}$ of the cylinder in the small tension, moderate tension and fixed length regimes.
\label{tab:R_chi_U_approx}}
\end{table}

\section{Energy comparison with other states}\label{sec:energy_comparison}

In the previous section, we have determined the parameters of the cylinder configuration and its energy in various cases. 
In this section, we compare the energy of the cylinder to the energy of other configurations:
\begin{itemize}
\item At low tension, $T\ll t^2$, the cylinder is compared to the helicoid.
\item At moderate tension but below the $\lambda$-point~\cite{Chopin2015}, $t^2\ll T\ll t$, the cylinder is compared to facets with isometric ridges~\cite{PhamDinh2016}.
\item Above the $\lambda$-point, $T\gg t$, or at fixed length, $\chi=0$,  the cylinder is again compared to the helicoid.
\end{itemize}

Comparing the result of the energies balance to transitions found by a linear stability analysis allows us to discuss the nature, continuous or discontinuous, of the transitions.

\subsection{Cylinder vs. helicoid at low tension}\label{}

At low tension, $T\ll t^2$, a linear stability analysis has been used to show that the helicoid is stable for $\eta\lesssim 10t$~\cite{Green1936, Green1937, Chopin2013, Chopin2015}.
Here, we compare the energy of the helicoid to the energy of the cylinder.
In this regime, the energy of the helicoid is given by~\cite{Chopin2015}
\begin{equation}
U_\text{hel} = \frac{\eta^4}{1440},
\end{equation}
and the energy of the cylinder is given by $U_\text{cyl} \simeq \eta^2 t^2/24$, Eq.~(\ref{eq:energy_cyl_low_T}).
The cylinder is thus favorable for
\begin{equation}
\eta>\eta_\text{hel-cyl}=2\sqrt{15}t\simeq 8t.
\end{equation}
This transition is represented as a thick dotted line in Fig.~\ref{fig:phase_diag}(b). 
Unfortunately, it occurs at a tension that is too low to allow a comparison with experiments.

First, we note that both approaches give the same scaling, $\eta\sim t$, for the transition.
Second, the energy comparison shows that the energy of the cylinder is lower than the energy of the helicoid when the helicoid becomes linearly unstable, which suggests a discontinuous transition between the helicoid and the cylinder.

\subsection{Cylinder vs. Facets with Isometric Ridges}\label{}

Upon increasing the twist at moderate tension, $t^2\ll T\ll t$, facetted morphologies appear soon after the longitudinal instability of the helicoid~\cite{Chopin2013, Bohr2013, PhamDinh2016}. 
Depending on tension and twist angle, two distinct facetted shapes can be observed. They are discriminated by the nature of the ridges between adjacent facets, isometric or minimal~\cite{PhamDinh2016}.
The shape that prevails at small tension and large twist is the configuration where the facets are separated by isometric ridges (FIR).
As shown in \cite{PhamDinh2016}, the width of the FIR ridges is given by $w_r\sim \phi R_c$, where $\phi\sim\eta$ is the angle between two facets and $R_c\sim t/(\eta\sqrt{T})$ is the radius of curvature of the ridges, hence $w_r\sim t/\sqrt{T}$.
Facets can be observed as long as the width of the ridges remains small compared to the width of the ribbon, {\it i.e.}, $w_r\sim t/\sqrt{T}\ll 1$, which means that $T\gg t^2$. 
Since the isometric ridges are portions of cylinders, there is no way to distinguish the FIR from the cylinder shape introduced here, they should be identical when $T\ll t^2$.

In contrast, the cylinder and the FIR becomes really different configurations when $t^2\ll T\ll t$, allowing to compare their energies.
For the cylinder, we use the moderate tension estimate, Eq.~(\ref{eq:energy_moderate_tension}); since $T\ll t$, the first term dominates: $U_\text{cyl}\sim t\eta\sqrt{T}$.
The energy of the FIR is given by $U_\text{FIR}\sim t\eta^2\sqrt{T}+\eta^2 T$~\cite{PhamDinh2016}; since $T\gg t^2$, the second term dominates: $U_\text{FIR}\sim \eta^2 T$.
Finally, we get that the cylinder has a lower energy for
\begin{equation}\label{eq:eta_fir_cyl}
\eta \gtrsim \eta_\text{FIR-cyl}=\frac{t}{\sqrt{T}}.
\end{equation}
The line separating the FIR and the cylinder ends at $T\sim t^2$, $\eta\sim 1$, which is reminiscent of the a liquid-vapor critical point.

The theoretical phase diagram at low tension is represented in Fig.~\ref{fig:phase_diag}(b), and the scaling law for the FIR-cylinder transition, Eq.~(\ref{eq:eta_fir_cyl}), is compared to experiments (Fig.~\ref{fig:phase_diag}); a good agreement is found.

\subsection{Cylinder vs. helicoid at large tension}\label{sub:cyl_hel_large_tension}

When the tension exceeds the $\lambda$-point tension $T_\lambda\sim t$, the helicoid undergoes a transverse buckling instability upon increasing the twist~\cite{Chopin2013, Chopin2015}.
The critical twist can be determined numerically; at large tension, it behaves asymptotically as $\eta_\text{tr}\simeq 4.4t/\sqrt{T}$.
Here, we compare the energy of the helicoid and the energy of the cylinder; this is all the more tempting since the shape of the unstable mode of the helicoid points to a cylinder shape (Fig.~11(a) in Ref.~\cite{Chopin2015}).

The energy of the helicoid is given by~\cite{Chopin2015}
\begin{equation}\label{eq:energy_hel}
U_\text{hel} = \frac{\eta^4}{1440}+\frac{\eta^2 T}{24}-\frac{T^2}{2},
	\end{equation}
and we use the moderate tension estimate for the energy of the cylinder, Eq.~(\ref{eq:energy_moderate_tension});
note that the last terms in Eqs.~(\ref{eq:energy_moderate_tension}, \ref{eq:energy_hel}) are identical.
The critical tension for a given twist can be determined exactly (App.~\ref{ap:hel_cyl_large_tension});
it is compared to experiments in Fig.~\ref{fig:phase_diag}(a) and a quantitative agreement is found.
 
At large tension and small twist, when $T\gg\eta^2$, the second term is larger than the first in the energy of the helicoid, Eq.~(\ref{eq:energy_hel}). The cylinder is energetically preferred when 
\begin{equation}\label{eq:eta_hel_cyl_large_tension}
\eta > \eta_\text{hel-cyl} = 4\sqrt{3}\frac{t}{\sqrt{T}}\simeq 6.9 \frac{t}{\sqrt{T}}.
\end{equation}
As in the very low tension regime, we find that the linear stability analysis and the comparison of energies give the same scaling law for the critical twist.
Here, the energy of the helicoid is smaller than the energy of the cylinder when the helicoid becomes linearly unstable, which points to a continuous transition between the two states.

\subsection{Cylinder vs. helicoid at constant length}\label{}

Finally, we discuss the helicoid-cylinder transition at constant length, which is particularly convenient to investigate in experiments.
This is a particular case of the helicoid-cylinder transition at large tension presented in the previous section.
We restrict ourselves to the case where the length is fixed to the rest length of the ribbon, $\chi=0$.
For the helicoid the tension is then given by $T_\text{hel}=\eta^2/24$~\cite{Chopin2013, Chopin2015}, while for the cylinder $T_\text{cyl}\simeq 0.28(\eta t)^{2/3}$ (Eq.~(\ref{eq:tension_const_length})).
The transition between the helicoid and the cylinder appears when the tension is measured while the twist is increased (Fig.~\ref{fig:eta_helcyl_chi0}(a)). 

The energy of the helicoid is obtained by using $T=\eta^2/24$ in Eq.~(\ref{eq:energy_hel}), leading to 
\begin{equation}
U_\text{hel}^{\chi=0} = \frac{\eta^4}{640},
\end{equation}
and the energy of the cylinder is given in Eq.~(\ref{eq:cyl_energy_chi0}).
The energy of the cylinder is lower for
\begin{equation}
\eta>\eta_{\mathrm{hel-cyl}}^{\chi=0}=\frac{240^{3/8}}{6^{1/4}}\sqrt{t}\simeq 5.0\sqrt{t}.
\end{equation}
For comparison, the linear stability analysis of the helicoid predicts that it buckles in the transverse direction at $\eta_\text{tr}\simeq 3.7\sqrt{t}$ (App.~\ref{ap:lin_stab_fixed_length}~\cite{Chopin2015}).
As in the previous paragraph, the helicoid becomes linearly unstable before while its energy is still lower than the energy of the cylinder.

As shown in Fig.~\ref{fig:eta_helcyl_chi0}(b), the twist angle at the helicoid-cylinder transition under constant length measured for various ribbon thicknesses are in quantitative agreement with the theoretical prediction.

\section{Conclusion}\label{sec:conclusion}

We have proposed a simple ansatz for the shape of a stretched and twisted ribbon at large twist, whereby the ribbon ``wraps'' around a cylinder. 
By comparing the energy of this shape to the energy of previously identified shapes, helicoid and facetted, we have shown that the cylinder is predicted to appear in a large twist region of the phase diagram, which is in good agreement with the experimental observations.
Moreover, the scaling laws of the different transitions are the same as those obtained by linear stability analysis.
Our analysis thus pushes forward our understanding of the phase diagram of the stretched and twisted ribbon by adding to the helicoid another well characterized shape.

The helicoid served as a basis to understand more complex morphologies because its stress field was easily and completely characterized.
In particular, this knowledge allows one to use linear stability analysis to determine the region where it should be observed.
A complete mechanical analysis of the cylinder remains however to be done, and at this stage we do not even know if the cylinder is a solution of the Föppl-von K\'arm\'an equations of elastic sheets.

Beyond the characterization of the cylinder, a unifying framework to describe the different ribbon morphologies is still desirable. 
As yet, our understanding of these morphologies rests on the linear stability analysis of the helicoid and on various ansatz for the cylinder, but also for the facetted shapes~\cite{PhamDinh2016}.
A promising route is to represent the ribbon as a ruled surface, which is always possible in the inextensible case~\cite{Starostin2007b, Korte2011, Dias2014}.
In this case, the ribbon can be described through its center line, with the direction of the generatrices as an internal variable.
The main drawback of these approaches is that they cannot describe the helicoid, where the ribbon is stretched.
It should be noted however that the helicoid is still a ruled surface, as well as the facets with isometric ridges introduced in Ref.~\cite{PhamDinh2016}, and the cylinder introduced here.
Hence, if the ruled surface approach could be extended to the extensible case, a unified description of these three shapes would be within reach.

\begin{acknowledgments}
The authors would like to thank Benjamin Davidovitch for stimulating discussions.
This work was partially supported by a grant from the 344 Belgian CUD program
\end{acknowledgments}

\appendix

\section{Exact relation between the tension, the contraction, and the radius}\label{ap:tension_R_chi}

We start from Eqs.~(\ref{eq:strain}, \ref{eq:elastic_energy}):
\begin{align}
U^\text{el}_\text{cyl} &= \frac{\epsilon^2}{2}+\frac{t^2}{24 R^2},\\
\epsilon & = \sqrt{(1-\chi)^2+\eta^2 R^2}-1.
\end{align}

The tension is given by Eq.~(\ref{eq:def_tension}):
\begin{equation}
T = -\frac{\partial U^\text{el}_\text{cyl}}{\partial \chi}=-\epsilon \frac{\partial\epsilon}{\partial\chi}.
\end{equation}
The derivative of the strain with respect to contraction is 
\begin{equation}
\frac{\partial\epsilon}{\partial\chi} = \frac{\chi - 1}{\sqrt{(1-\chi)^2+\eta^2 R^2}} = \frac{\chi-1}{1+\epsilon},
\end{equation}
so that we have for the tension
\begin{equation}\label{eq:tension_epsilon_chi}
T = \frac{\epsilon}{1+\epsilon}(1-\chi).
\end{equation}

We can use that the radius minimizes the energy:
\begin{equation}
0=\frac{\partial U_\text{cyl}^\text{el}}{\partial R} = \epsilon \frac{\partial \epsilon}{\partial R}-\frac{t^2}{12 R^3}.
\end{equation}
The derivative of the strain with respect to the radius is
\begin{equation}
\frac{\partial \epsilon}{\partial R} = \frac{\eta^2 R}{\sqrt{(1-\chi)^2+\eta^2 R^2}} = \frac{\eta^2 R}{1+\epsilon}.
\end{equation}
Inserting this relation into the previous one, we get
\begin{equation}
\frac{\epsilon}{1+\epsilon} = \frac{t^2}{12\eta^2 R^4}.
\end{equation}
Using this equation in Eq.~(\ref{eq:tension_epsilon_chi}), we finally obtain
\begin{equation}
T = \frac{t^2(1-\chi)}{12\eta^2 R^4}
\end{equation}

\section{Critical tension for the helicoid-cylinder transition at large tension}\label{ap:hel_cyl_large_tension}

Here, we determine analytically the critical tension for the helicoid-cylinder transition at large tension (Sec.~\ref{sub:cyl_hel_large_tension}).
Comparing the energies of the cylinder (Eq.~(\ref{eq:energy_moderate_tension})) and of the helicoid (Eq.~(\ref{eq:energy_hel})) leads to a transition when
\begin{equation}
\frac{\eta^4}{1440}+\frac{\eta^2 T}{24}= \frac{t\eta\sqrt{T}}{2\sqrt{3}}.
\end{equation}
This equation can be solved for the tension $T$:
\begin{equation}\label{eq:hel_cyl_analytical tension}
T=\frac{12 t^2}{\eta^2}\left(1\pm\sqrt{1-\frac{\eta^4}{720 t^2}} \right)^2.
\end{equation}
The large tension regime corresponds to the $+$ sign.

The asymptotic law at large tension can be recovered by neglecting the term $\eta^4/(720t^2)$ in the square root; we get $T=48t^2/\eta^2$, which is equivalent to Eq.~(\ref{eq:eta_hel_cyl_large_tension}).
This approximation is valid if
\begin{equation}
1\gg\frac{\eta^4}{t^2}\sim \frac{t^2}{T^2},
\end{equation}
which is the case for $T\gg t$.

\section{Linear stability analysis of the helicoid at fixed length}\label{ap:lin_stab_fixed_length}

Imposing a zero contraction in the helicoid sets the tension to $T=\eta^2/24$ and the stress field to (Eqs.~(3, 4) of Ref.~\cite{Chopin2015})
\begin{align}
\sigma^{ss}_\mathrm{hel}(r) & = \frac{\eta^2 r^2}{2},\label{eq:longitudinal_stress}\\
\sigma^{rr}_\mathrm{hel}(r) & = \frac{\eta^4}{8}\left(r^2-\frac{1}{4} \right) \left(r^2+\frac{1}{4}\right),\label{eq:transverse_stress}
\end{align}
where $s$ is the longitudinal component and $r$ the transverse component.

We can apply the linear stability analysis of Ref.~\cite{Chopin2015} (Sec. 4.3) for the transverse buckling with this stress field, in the case of an infinitely long ribbon.
Inserting Eqs.~(\ref{eq:longitudinal_stress}, \ref{eq:transverse_stress}) in the buckling equation (Eq.~(79) of~\cite{Chopin2015}), we see that the critical value for $\eta$ scales as $\eta\sim\sqrt{t}$.
A numerical resolution of the buckling equation gives the numerical factor:
\begin{equation}\label{eq:etac_linstab}
\eta\simeq 3.7\sqrt{t}.
\end{equation}


\end{document}